\documentclass[usenatbib,usegraphicx,useAMS,usenatbib]{mn2e}

\newcommand{\water}{H$_2$O}

\newcommand{\kms}{km\,s$^{-1}$}

\title[High velocity \water~masers in OH~009.1--0.4]{High velocity \water~maser
emission from the Post-AGB star OH~009.1--0.4}
\author[Walsh et al.]{A. J. Walsh$^{1}$\thanks{E-mail:
andrew.walsh@jcu.edu.au}, S. L. Breen$^{2,3}$, I. Bains$^{4}$ and W. H. T. Vlemmings$^{5}$
\\
$^{1}$School of Maths, Physics and IT, James Cook University, Townsville, QLD 4811, Australia; \\
$^{2}$School of Mathematics and Physics, University of Tasmania, Private Bag 37, Hobart, TAS 7001, Australia; \\
$^{3}$Australia Telescope National Facility, CSIRO, PO Box 76, Epping, NSW 2121, Australia; \\
$^{4}$Centre for Astrophysics and Supercomputing, Swinburne University of Technology, PO Box 218, Hawthorn, VIC 3122, Australia;\\
$^{5}$Argelander Institute for Astronomy, University of Bonn, Auf dem H\"{u}gel 71, 53121 Bonn, Germany}
\begin{document}

%\date{Accepted 1988 December 15. Received 1988 December 14; in original form 1988 October 11}

%\pagerange{\pageref{firstpage}--\pageref{lastpage}} \pubyear{2002}

\maketitle

\label{firstpage}

\begin{abstract}
Observations of \water~masers towards the post-AGB star and water fountain source OH~009.1--0.4
were made as part of HOPS (The \water~southern galactic Plane Survey), with the Mopra radiotelescope.
Together with followup observations using the Australia Telescope Compact Array (ATCA), we have
identified \water~maser emission over a velocity spread of nearly 400\kms~(--109 to +289\,\kms).
This velocity spread appears to be the largest of any known maser source in our Galaxy.
High resolution observations with the ATCA indicate the maser emission is confined to a
region $0\farcs3 \times 0\farcs3$ and shows weak evidence for a separation of the red-
and blueshifted maser spots. We are unable to determine if the water fountain is projected along
the line of sight, or is inclined, but either way OH~009.1--0.4 is an interesting source,
worthy of followup observations. 
\end{abstract}

\begin{keywords}
masers -- stars: AGB and post-AGB
\end{keywords}

\section{Introduction}

Water (\water) maser emission is found throughout our Galaxy,
pinpointing unusual physical conditions towards many different
astrophysical objects, including high- and low-mass star forming regions, 
planetary nebulae \citep{miranda01}, Mira variables \citep{hinkle79}
and asymptotic giant branch (AGB) stars \citep{engels96}.
They are also found outside our Galaxy,
such as in the centres of active galaxies \citep{claussen84}.
In order to study the occurrence of \water~masers in our Galaxy, a
large-scale untargetted survey has begun, called HOPS (The \water~southern
galactic Plane Survey; Walsh et al. {\em in preparation}).
This survey will cover 90 square
degrees of the southern Galactic plane, searching for, amongst other things,
\water~masers.\\

Planetary nebulae (PN) are the end-phase of evolution for intermediate-mass
($\sim$~1-8 M$_\odot$) Main Sequence stars. These stars spend a significant
fraction of their lives ($\sim$~10$^6$~yr) on the Asymptotic Giant Branch
(AGB), a phase characterised by copious mass-loss which results in thick,
dusty, molecular envelopes (see e.g. Herwig 2005 for a review of AGB
evolution) with expansion velocities of order ~10-20~km~s$^{-1}$. In the
O-rich case, the envelopes of these AGB stars can display emission from the
masing species of SiO, OH and H$_2$O. The velocity profiles of these species
(e.g. Deacon et al. 2004, 2007) are usually consistent with them having arisen
in largely spherical shells around their host star (e.g. Cohen 1989);
additionally, the velocity extent of the OH typically lies outside of and
encloses that of the H$_2$O. Interferometric observations of these species
reveal spatiokinematic distributions of maser spots that largely corroborate
the suggested overall sphericity, sometimes with small deviations within the
inner molecular envelopes (e.g. Bains et al. 2003). As the objects evolve past
the tip of the AGB, it is surmised that a fast wind emanates from the
exposed stellar core \citep{kwok78} and gradually the maser species are extinguished due to
the loss of coherence length in the disrupted shell (see e.g. van Winckel
2003 for a review of post-AGB evolution). It is this fast wind that is
thought to amplify any existing asymmetries in the 
slower AGB envelope, leading to the morphological evolution of the object 
from largely spherical to the array of shapes displayed by PN (e.g. Balick \& 
Frank 2002).\\

%PN are known to display complex morphologies that range from axisymmetric
%structures (viz. (multi-) polar lobes, outflowing point-symmetric knots,
%pinched waists) through spherical and ring-like shells to completely
%asymmetric features. The mechanism by which the largely spherical AGB
%envelope is thought to be transformed into the array of shapes shown by PN
%is a matter of contention, with binarity or magnetic fields, or a
%combination of the two, being proposed as the likely contenders to form a
%latitude-dependent density enhancement that acts to sculpt the outflowing
%fast wind. One of the reasons this mechanism remains unclear is because the
%transition phase between AGB and PN (the post-AGB or pAGB phase) is
%rapid---of order 1000s of years---and during this time the objects are
%largely obscured by their dust shells; their study, therefore, is inhibited.
%\\
%
Recent imaging of post-AGB dust shells (e.g. Gledhill 2005) in scattered light has
revealed a preponderance of axisymmetric structures, suggesting the window
for turning on the shaping mechanism can be narrowed to the late AGB/early
post-AGB phase. Further, the slowly-growing subclass of `water fountain' sources
(first discovered and aptly named by Likkel \& Morris 1988), with about 10
members \citep{imai07}, can potentially elucidate the mystery of the shaping
mechanism. Water fountains are bipolar post-AGB stars that display highly
collimated and high velocity expanding regions (typically $v_{\rm
exp}$~$<$~100~km~s$^{-1}$) of \water~maser emission. The spectral profile
of the \water~maser emission displays a larger velocity range of emission compared
to that of OH, where present; maps of the sources reveal that the OH emission traces
the remnant molecular shell while the \water~emission is shock-excited in
collimated bipolar jets \citep{imai02}. Observations seem to confirm the
theory that these high velocity jets are likely driven by
magneto-hydrodynamical processes \citep{vlemmings06}. However, whether
the observed strong magnetic fields originate from single stars or through
binary interactions is still unclear. Water fountains may therefore represent an
extremely short-lived phase of post-AGB evolution that all such objects pass
through, where much of the shaping takes place via the sculpting caused by
the high velocity jets. Alternatively, they may be a special subset of
stars, perhaps the more massive ones that evolve quicker and have
sufficiently thick envelopes to shield clouds of \water~vapour and preserve
them to later be shocked into masing by the fast wind (Suarez et al. 2007).
Either way, their discovery represents an exciting new addition to the study
of the late stages of stellar evolution.\\

In this Letter, we report on the detection of a \water~maser towards the
young post-AGB star known as OH~009.1--0.4 and b292 \citep{sevenster97},
which shows maser emission over the widest velocity
range known in our Galaxy, confirming it as a new addition to the short list of
known water fountains. The post-AGB star is associated with the IRAS
source 18043--2116. It is unusual in that it exhibits 1665\,MHz and 1612\,MHz
OH maser emission, but not at 1667\,MHz \citep{deacon04}. Furthermore,
it is the only known post-AGB star to show 1720\,MHz OH maser emission,
which is thought to arise in the region where the AGB and starting planetary
nebula winds collide \citep{sevenster01}. Previous observations of
\water~maser emission towards OH~009.1--0.4 \citep{deacon07} showed emission
covering 210\,\kms, although the authors noted that there could be other
maser features that were outside the observed velocity range.\\

\section{Observations}
\subsection{Mopra Observations}

The Australia Telescope National Facility Mopra telescope is a 22\,m antenna
located 26\,km outside the town of Coonabarrabran in New South Wales, Australia.
It is at an elevation of 850 metres above sea level and at a latitude of
31$^\circ$ south.\\

Observations of OH~009.1--0.4 were performed as part of the large survey HOPS
on 2008 March 9 and 10, using the on-the-fly mapping method. Total on-source
integration time is approximately two minutes. The Mopra spectrometer (MOPS)
was used in ``zoom'' mode, which allows us to simultaneously observe 16
different frequencies, each with 4096 channels across a bandwidth of
137.5\,MHz, corresponding to 0.45\,\kms~per channel. To observe \water~masers,
we employed two contiguous zoom bands, allowing us to search for maser emission
over a velocity range between --2395 and +1326\,\kms.\\

Data were reduced using
livedata and gridzilla, which are both AIPS++ packages written for the
Parkes radiotelescope and adapted for Mopra.

\subsection{ATCA Observations}

Followup observations of OH~009.1--0.4 were made with the Australia Telescope
Compact Array (ATCA) on 2008 July 12 and 13, using the H214 hybrid array.
Observations were made using three different frequencies spaced 26\,MHz
apart, each 32\,MHz wide and containing 512 channels, equivalent to
0.84\,\kms~per channel. Together, the three frequencies covered a velocity
range of --395 to +670\,\kms with respect to the local standard of rest (LSR).
Observations were made with one linear polarisation,
were phase referenced to either 1730--130 or 1817--254 and flux calibrated
using 1934--638. We estimate the flux scale will be accurate to around 10\%,
which is typical for observations at this frequency.\\

For each frequency, seven snapshot observations of two minutes each were taken.
The snapshots were evenly spaced throughout eight hours, to ensure a wide
coverage of the UV plane. The data were reduced using the {\sc miriad} package
using similar techniques described in \citep{walsh07}. Strong maser emission
was used to self-calibrate the data, and allowed data from antenna 6 to be
included, greatly reducing the synthesised beam to 1.19\arcsec $\times$
2.75\arcsec.

\section{Results}

\begin{figure}
\includegraphics[width=0.5\textwidth]{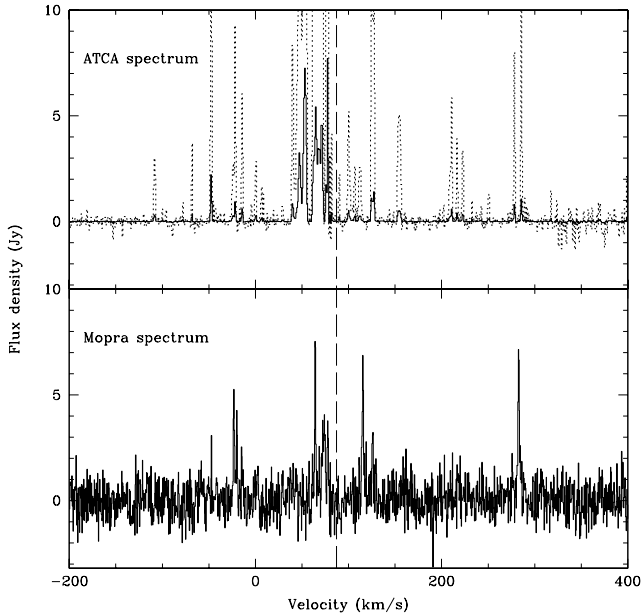}
\caption{\water~maser spectra for OH~009.1--0.4. The bottom spectrum was
taken using the Mopra radiotelescope. The top spectrum is combined data
from three bandpasses using the ATCA and was taken approximately 125 days
after the Mopra spectrum. The vertical dashed line represents the systemic velocity
of 87\,\kms~\citep{deacon07}. The dotted lines in the upper (ATCA) spectrum
show 10$\times$ the solid line.}
\label{fig1}
\end{figure}

The initial Mopra observations showed maser emission detected at velocities
from --48 to +286\,\kms, as shown in the bottom spectrum of Figure \ref{fig1}.
This emission comprises of eleven individual maser spots. Followup observations
with the ATCA (top spectrum in Figure \ref{fig1}) showed emission spanning a
larger velocity range from --109 to +289\,\kms~and comprising of 32
individual maser spots. The greater number of maser spots identified in the
ATCA observations appears to be a result of the lower noise level
in the ATCA spectrum, allowing weaker spots to be identified.
However, intrinsic variablility of the masers has resulted in some brighter spots in
the ATCA observations. This variability has taken place over 125 days and is
most noticeable in the velocity range of +40 to +80\,\kms, where the brightest
emission occurs in the ATCA spectrum.\\

Some overlap in the ATCA bandpasses allowed us to observe the same
maser spots between --50 and --12\,\kms~on different nights. Based on the two
brightest maser spots ($>$1\,Jy) within this range (at --21.8 and
--47.7\,\kms), we find an absolute offset of not more than 1.9\arcsec~between
the observations made on different nights. We note that the weather on the second
night of observations was considerably worse than the first night, with significantly
higher phase noise. The second nights
observations covered maser spots in the velocity range --12\,\kms~to --109\,\kms.
Thus, we expect absolute positional data for the majority of maser spots,
observed on the first night, to be better than
1.9\arcsec. We use the brightest maser spot in the spectrum
(based on integrated intensity, rather than the peak) as the reference
position of our observations, which is at 18$^h$07$^m$20\,$\fs$85
--21$^\circ$16$\arcmin$12$\farcs$0 (J2000).\\

With the small ATCA synthesised beam, we are able to accurately determine
relative positions of each maser spot. For each maser spot, we integrated all
channels showing emission, avoiding any channels which might be confused by
emission from other spots. The resultant integrated image was fitted with a
two-dimensional gaussian, using the ``imfit'' routine in {\sc miriad}. For
each maser spot, this allows us to determine its relative position more
accurately than the beam size. We estimate the relative positional uncertainty
of each maser spot as $\Theta_{\rm beam}/2{\rm SNR}$ \citep{fomalont99},
where $\Theta_{\rm beam}$ is the synthesised beam (1.19\arcsec $\times$
2.75\arcsec) and SNR is the signal-to-noise ratio.\\

\begin{figure}
\includegraphics[width=0.5\textwidth]{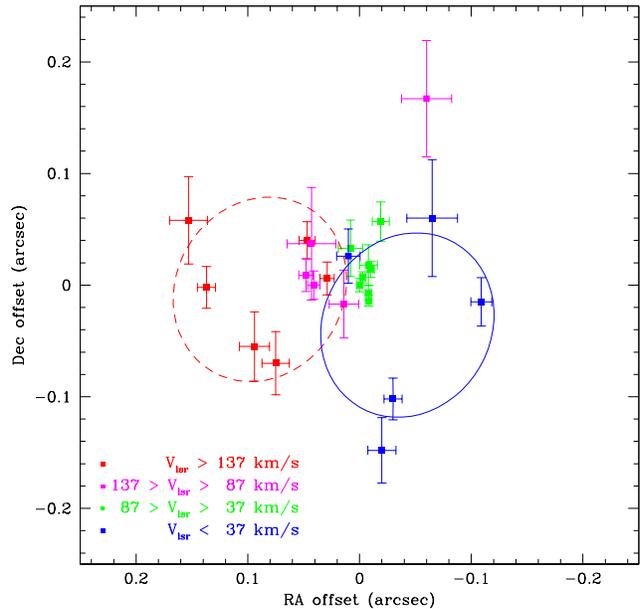}
\caption{Relative positions of \water~maser spots in OH~009.1--0.4.
Each maser spot is coloured according to their line-of-sight velocity. The error
bars on each maser spot represent the estimated relative positional uncertainty
based on the synthesised beam width and the signal-to-noise ratio. Positions
are given relative to the strongest maser spot, which is located at
18$^h$07$^m$20\,$\fs$85 --21$^\circ$16$\arcmin$12$\farcs$0 (J2000) and occurs
at a velocity of 53.1\,\kms. Red and blue ellipses indicate the offset and extent
of the high velocity red- and blue-shifted components.}
\label{fig2}
\end{figure}

\section{Discussion}

We have detected \water~maser emission over a radial velocity range of
398\,\kms~which we believe is the largest of any known water fountain,
or indeed any maser source in our Galaxy. Based on the available information,
we are unable to determine the orientation of this system
with respect to the line of sight. If the orientation is not
along the line of sight, the spread of maser velocities will be even greater.
For example, if the orientation is 45$^\circ$ to the line of sight, the masers
will cover a velocity range of over 560\,\kms~in three dimensional space.
This can be compared to recent observations of the water fountain
IRAS 16342--3814 \citep{claussen08}, whose radial velocities
are measured over a spread of 270\,\kms~and whose three dimensional
velocities range up to 370\,\kms, based on an inclination angle
of 45$^\circ$. Thus, if OH~009.1--0.4 has a large inclination angle, it shows
maser velocities far in excess of any other known water fountain. Alternately,
it may be aligned close to the line of sight with a three dimensional velocity
slightly larger than that of IRAS 16342--3814.

Based on the systemic velocity of the post-AGB star of
87\,\kms~\citep{deacon07}, the highest velocity spots of the \water~maser
spectrum are at --196 and +202\,\kms~along the line of sight, relative to the
systemic velocity.
It is interesting to note that the velocity difference of the most redshifted
maser spot to the systemic velocity closely matches the magnitude
of the velocity difference between the most blueshifted component and the
systemic velocity. In addition to this, we find 8 out of 13 maser spots (60\%)
on the redshifted side have maser spot counterparts on the blueshifted side
with similar magnitudes of velocity offset from the systemic velocity.
Such mirroring of the maser emission could be a result of symmetries
between each lobe of a bipolar jet.\\

Figure \ref{fig2} shows the relative positions of the brightest 24 maser spots,
together with their error bars. As can be seen from this Figure, all the
emission is found within a small area of $0\farcs3 \times 0\farcs3$. However,
we do see significant differences between the positions of some maser spots.
We assume a kinematic distance to OH~009.1--0.4 of 6.4\,kpc \citep{brand93}. Thus,
the maser spots are confined to a projected area of $1900 \times 1900$\,AU.

As seen in Figure \ref{fig2}, the H$_2$O maser morphology of OH~009.1--0.4
does not resemble the distinct bipolar structures characteristic of
the imaged water fountain sources. Although there is a offset of
$\sim0.13\arcsec$ (corresponding to $\sim 830$~AU) between the most
red- and blue-shifted maser features, the opening angle of the outflow
of OH~009.1--0.4 appears to be much larger than that of the collimated
jet of the typical water fountain sources. However, this could be due
to projection effect when the source is observed almost along the
outflow (`jet') axis. Assuming a regular bipolar outflow structure,
the opening angle of the OH~009.1--0.4 outflow is given by $\tan(\phi)
\approx 2.7\sin(i)$, with $i$ the inclination angle of the outflow
axis from the line-of-sight. Ignoring any possible contributions from
jet precession as observed for W43A \citep{imai02}, a typical water
fountain opening angle $\phi<10^\circ$ implies $i \leq 4^\circ$. We
can also use the projected separation to estimate the dynamical age of
the high velocity outflow from OH~009.1--0.4. Again assuming a regular
bipolar outflow with constant velocity, the dynamical age
$t_{\rm outf}\approx 10/\tan(i)$~yr. While a dynamical age as low as 10~yr
cannot be excluded, a more typical water fountain age of $\sim
50-100$~yr \citep{imai07} also implies a small inclination
angle. Thus, the high velocity H$_2$O masers around OH~009.1--0.4 either
originate from a typical water fountain jet observed almost along the
jet axis, or they occur in an extremely young and fast wide-angled
wind. These different scenarios can be directly tested by astrometric
monitoring at Very Long Baseline Interferometry (VLBI) resolution, as
a regular outflow implies proper motions of
$\mu\sim13\tan(i)$~mas~yr$^{-1}$ for the high velocity H$_2$O maser
components.

Contrary to the other water fountain sources, the strongest H$_2$O
maser emission of OH~009.1--0.4 originates from within $\sim 50$\,\kms~of
the stellar velocity. While the spatial extent and velocity of these
masers are fairly typical for the standard AGB H$_2$O maser shells
\citep[e.g.][]{bains2003}, Figure \ref{fig2} shows that the low velocity
emission at the blue- and red-shifted sides of the spectrum also
arises in separated regions. Although high resolution observations are
needed to properly probe the relation between the high- and
low-velocity components, the emission could arise in an
$\sim50$\,\kms~equatorial wind as observed at somewhat lower velocities
($\sim30$\,\kms) in the water fountain sources IRAS~18286--0959 and
IRAS~18460--0151 \citep{deguchi08, imai07}. Alternatively, the
low-velocity components are excited in a slow wind or in accelerating
material further down the fast outflow, possibly due to episodic
ejections.

%It appears from Figure \ref{fig2} that the maser spots showing the most
%redshifted and blueshifted velocities (red and blue, respectively, in
%Figure \ref{fig2}) tend to occur in separated positions on the sky, with
%those maser spots close to the systemic velocity (green and orange) in between.
%This is suggestive of a bipolar structure commonly seen in other water fountain
%sources, but the distribution in Figure \ref{fig2} does not show a clear
%collimated structure, nor are the red and blueshifted parts well separated.
%Further observations at higher resolution will be required to properly identify the
%structure and orientation of any bipolar lobes.

\section{Conclusions}
We have observed the water fountain OH~009.1--0.4 with both the Mopra and
ATCA radiotelescopes. We find water maser spots covering an unprecedented
velocity range of 398\,\kms~which we believe is the largest range of velocities
for any known maser site in our Galaxy. Using the ATCA interferometer, we have partially
resolved the water maser emission, which covers a projected area of
$1900 \times 1900$\,AU. From our current observations, it is not clear if the
water fountain is aligned along the line of sight, or if it is inclined. If it
is aligned along the line of sight, then this orientation offers a unique perspective
on water fountains with the jet pointed towards us. If it is inclined to the line of
sight, then the three dimensional velocities will be even larger than measured here.
Either way, OH~009.1--0.4 is certainly a unique and interesting water fountain,
worthy of further investigation.\\

We suggest that followup observations OH~009.1--0.4 at higher spatial resolution,
such as with VLBI, will be able to tell us more about the nature of this region.\\

%\section*{Acknowledgments}

%\bsp

\label{lastpage}

\end{document}